\documentclass{elsart}

\usepackage{graphics}
\usepackage{epsfig}

\usepackage{amssymb}

\begin{document}

\begin{frontmatter}



\title{Coherent Multiple Scattering Effect in DIS\thanksref{grant}}
\thanks[grant]{Supported in part by the National Science Fundation, 
grant PHY-0340729.}


\author{Xiaofeng Guo and Jun Li}

\address{Department of Physics and Astronomy, Iowa State University \\
         Ames, Iowa 50011, U.S.A.}

\begin{abstract}
We present a QCD calculation of the coherent multiple scattering effect 
on single hadron production in semi-inclusive lepton-nucleus 
deep inelastic scattering (DIS). We show that the destructive 
interference of coherent multiple scattering leads to suppression 
in hadron productions. The suppression is similar 
to the expectations of energy loss calculations. 
We compare our calculation with experimental data.
\end{abstract}

\end{frontmatter}

\noindent{\it 1.\ Introduction}\
The strong suppression of high transverse momentum hadron in 
relativistic heavy ion collisons was considered to be 
an evidence for the  QCD quark-gluon plasma \cite{rhic-pion}.
The suppression was believed to be the result of 
medium induced 
radiative energy loss of high energy partons \cite{jet-quenching}.
However, recent data indicate that heavy quarks would have to lose 
the same amount of energy as that of a light quark  
if the radiative energy loss is the only source of the suppression 
\cite{charm-lose}.  On the other hand, we expect heavy quarks to lose
much less energy than a light quark because of its mass \cite{dead-cone}.  
In this talk, we present a new source of suppresion from 
the destructive multiple scattering in a nuclear medium \cite{guo-li}.  
We calculated the single hadron production 
in DIS in perturbative QCD, and  
found that the coherent interference results into a suppression 
in the hadron production rate.  

\noindent{\it 2.\ Single hard scattering}\
We consider the production of a single hadron of momentum $p_h$ in
lepton-nucleus deep inelastic scattering (DIS),
$e(k_1) + A(p) \longrightarrow e(k_2) + h(p_h) +X$, with the target
atomic weight $A$ and average nucleon momentum $p=P_A/A$.
The semi-inclusive DIS cross section, 
\begin{equation}
\frac{d{\sigma}_{eA\rightarrow e h X}}{dx_B dQ^2 dz_h}
=\frac{1}{8\pi}\, \frac{e^4}{x_B^2s^2Q^2} \,
L^{\mu\nu}(k_1,k_2)\, \frac{dW^A_{\mu\nu}}{dz_h}\ ,
\label{sigma-c}
\end{equation}
where $s=(p+k_1)^2$ is the total invariant mass of the lepton-nucleon
system, and 
$z_h \equiv {p\cdot p_h }/{p\cdot q} = {2x_Bp\cdot p_h}/{Q^2}$ 
is the fraction of the photon momentum carried by the observed hadron.
The Bjorken $x_B=Q^2/(2p\cdot q)$ with $Q^2=-q^2$.  
The leptonic tensor $L^{\mu\nu}(k_1,k_2)
={\rm Tr}(\gamma \cdot k_1 \gamma^{\mu}
\gamma \cdot k_2 \gamma^{\nu})/2$.  For single scattering 
at the lowest order, the hadronic tensor can be factorized as
\begin{equation}
\frac{dW^A_{\mu\nu}}{dz_h}=\frac{1}{2}\, e^T_{\mu\nu} \sum_q \,e_q^2\,
                        A\,\phi^A_{q}(x_B,Q^2)\, D_q(z_h, Q^2) \, ,
\label{FT-0} 
\end{equation} 
where $e^T_{\mu\nu}=[p_\mu q_\nu + q_\mu p_\nu]/{p\cdot q}
   +p_\mu p_\nu({2x_B}/{p\cdot q})  - g_{\mu\nu}$, 
$\phi^A_{q}(x_B,Q^2)$ is nuclear quark distribution 
per nucleon of flavor $q$,  and $D_q(z_h, Q^2)$ is quark-to-hadron 
fragmentation function.
The nuclear dependence of Eq.~(2) 
and  Eq.~(\ref{sigma-c}) is 
limited to the nuclear dependence of $\phi^A_{q}(x_B,Q^2)$.

\noindent{\it 3.\ Coherent multiple scattering}\
Multiple scattering 
introduces the medium size enhanced, or the $A^{1/3}$-type, 
nuclear effects.  We adopt the leading pole approaximation
to extract the leading contributions  in powers of 
$A^{1/3}$\cite{LQS2}.  

\begin{figure}
\begin{center}
\includegraphics[width=2.3in]{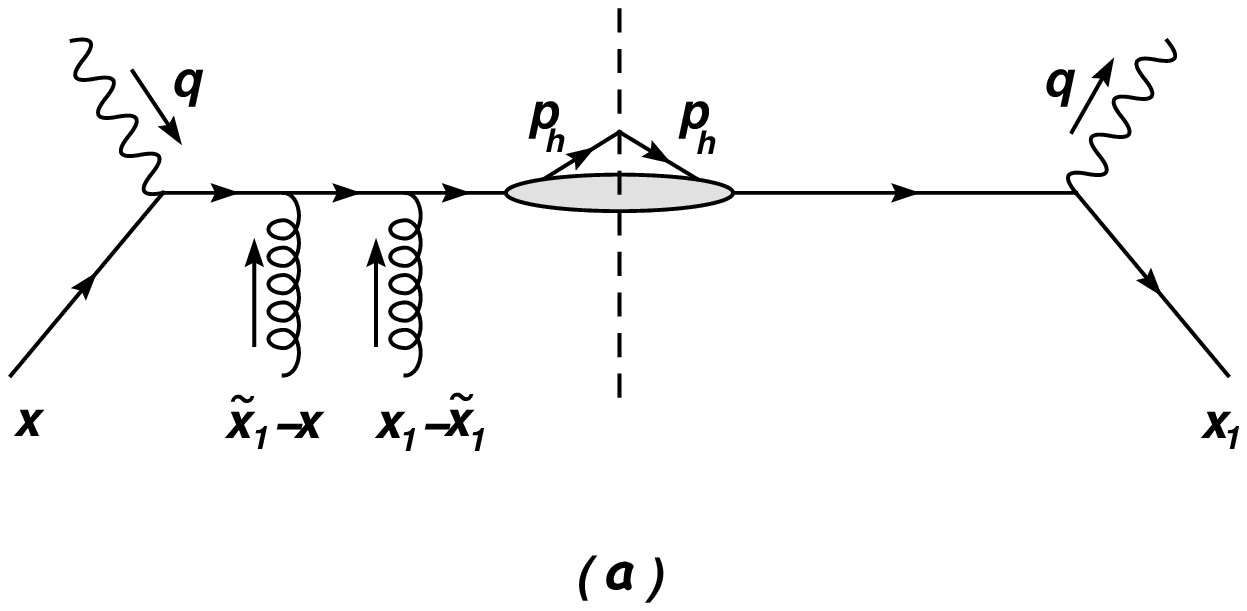}
\hskip 0.2in
\includegraphics[width=2.3in]{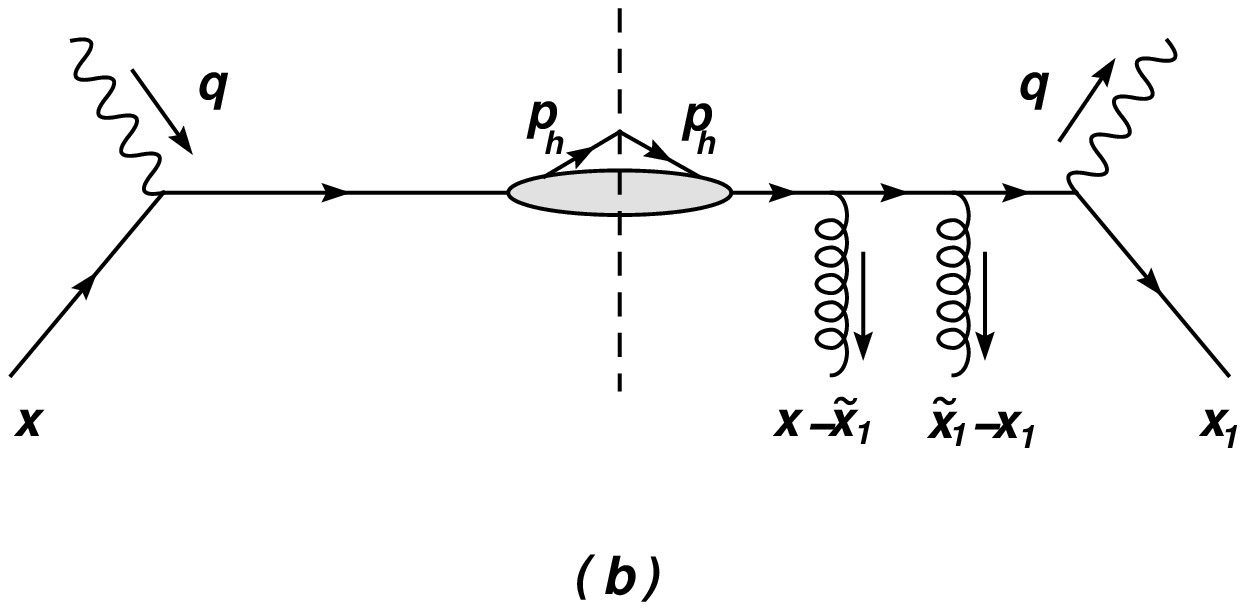}
\caption{Double scattering diagrams that contribute to the 
production rate of single hadron in DIS.}
\label{fig1}
\end{center}
\end{figure}

For double scattering at lowest order in $\alpha_s$, 
we only need to consider two diagrams in Fig.~\ref{fig1}.
When $x_B \ge x_c\sim 0.1$ \cite{qiu-vitev}, 
we  consider coherent multiple scattering effects 
between the scattered quark and the partons of the nuclear target 
at the same impact parameter in the Breit frame
and obtain \cite{guo-li} 
\begin{eqnarray}
\frac{dW_{\mu\nu}^{(1)}}{dz_h} 
& = & 
\frac{1}{2} e_{\mu\nu}^T \, \sum_q e_q^2
\left[\frac{4\pi^2\alpha_s}{3}\right]
    \frac{ z_h}{Q^2}\,T_{qg}^A(x_B,Q^2)\, \frac{dD(z_h, Q^2)}{dz_h}
\label{FT-D}
\\
&\approx &
\frac{1}{2} e_{\mu\nu}^T \, \sum_q e_q^2
\left(A\phi^A_{q}(x_B,Q^2)\right)
\left[\frac{z_h \xi^2 (A^{1/3}-1)}{Q^2}\right]
\frac{dD(z_h, Q^2)}{dz_h}
\label{FT-1}
\end{eqnarray}
where $\sum_q$ runs over all quark and antiquark flavors.
The twist-4 quark-gluon correlation function, $T_{qg}^{A}(x_B,Q^2)$, 
is defined in Ref.~\cite{LQS2}.  
In deriving Eq.~(\ref{FT-1}), we assumed that a large nucleus
is made of color neutral nucleons with a constant density. 
The quantity $\xi^2$ defined in Ref.~\cite{qiu-vitev} 
is proportional to a matrix element of a gluon density operator, and
represents a characteristic scale of quark interaction with 
nuclear medium.  
If $x_B < x_c$, the hadronic tensor in Eqs.~(\ref{FT-D}) and 
(\ref{FT-1}) has an additional term proportional to 
$x_B \frac{d}{dx_B}\phi_{q}(x_B,Q^2)$ \cite{guo-li}.

For $x_B \ge x_c$, we generalized our result in Eq.~(\ref{FT-1}) to
$n$ scattering \cite{guo-li}
\begin{eqnarray}
\frac{dW_{\mu\nu}^{(n)}}{dz_h}
&\approx & 
\frac{1}{2} e_{\mu\nu}^T \, \sum_q e_q^2
\left(A\phi^A_{q}(x_B,Q^2)\right)
\frac{1}{n!}
\left[\frac{z_h \xi^2 (A^{1/3}-1)}{Q^2}\right]^n
\frac{d^n D(z_h, Q^2)}{dz_h^n} \, .
\label{FT-n}
\end{eqnarray}
 We sum the $A^{1/3}$-type contributions in Eq.~(\ref{FT-n}) to 
$N$ scatterings, and take
$N\approx \infty$ because the effective value of $\xi^2$ is
relatively small. We obtain
\begin{eqnarray} 
\frac{dW^A_{\mu\nu}}{dz_h}  
& \approx & 
\frac{1}{2} e_{\mu\nu}^T \, \sum_q e_q^2
\left(A\phi^A_{q}(x_B,Q^2)\right)
\sum_{n=0}^{N}  \frac{1}{n!}
\left[\frac{z_h \xi^2 (A^{1/3}-1)}{Q^2}\right]^n
\frac{d^n D(z_h, Q^2)}{dz_h^n}
\nonumber \\
&\approx &
A \, 
\frac{1}{2} e_{\mu\nu}^T \, \sum_q e_q^2\,
\phi^A_{q}(x_B,Q^2)\,
D_q\left( z_h+ \frac{z_h \xi^2 ( A^{1/3}-1) }{Q^2}, Q^2 \right) \, .
\label{FT-res}  
\end{eqnarray}
Eq.~(\ref{FT-res}) is our main result.  
The net effect of coherent multiple scattering of 
a propagating quark in the medium, 
is equivalent to a shift, $\Delta z$, in the variable $z_h$ 
of the quark fragmentation function $D_q(z_h,Q^2)$ with
$
\Delta z = z_h\, \xi^2 ( A^{1/3}-1)/Q^2 ,
$
which leads to a suppression of the production rate
\cite{guo-li}.

\begin{figure}
\begin{center}
\includegraphics[width=2.5in]{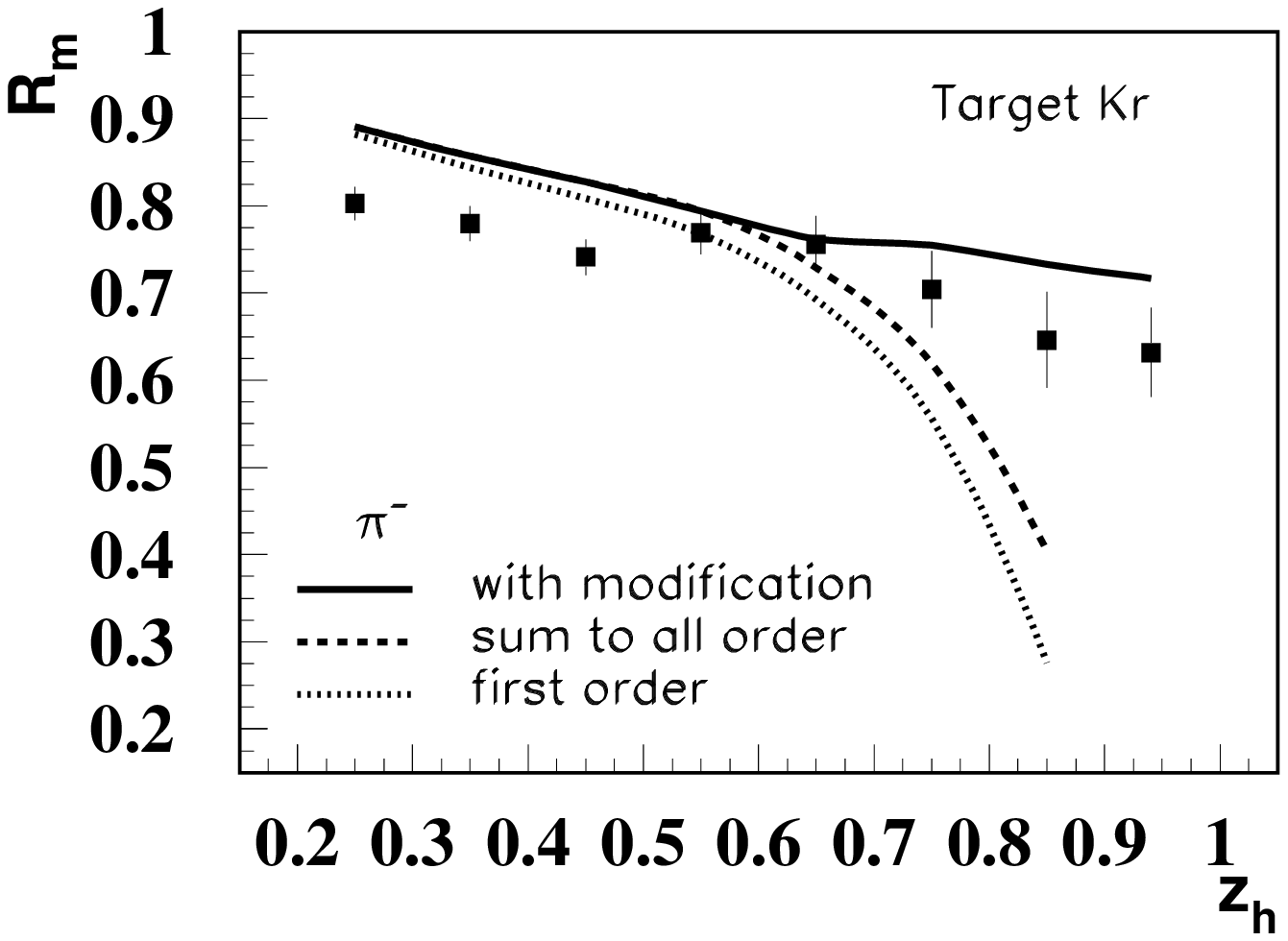}
\hskip 0.2in
\includegraphics[width=2.5in]{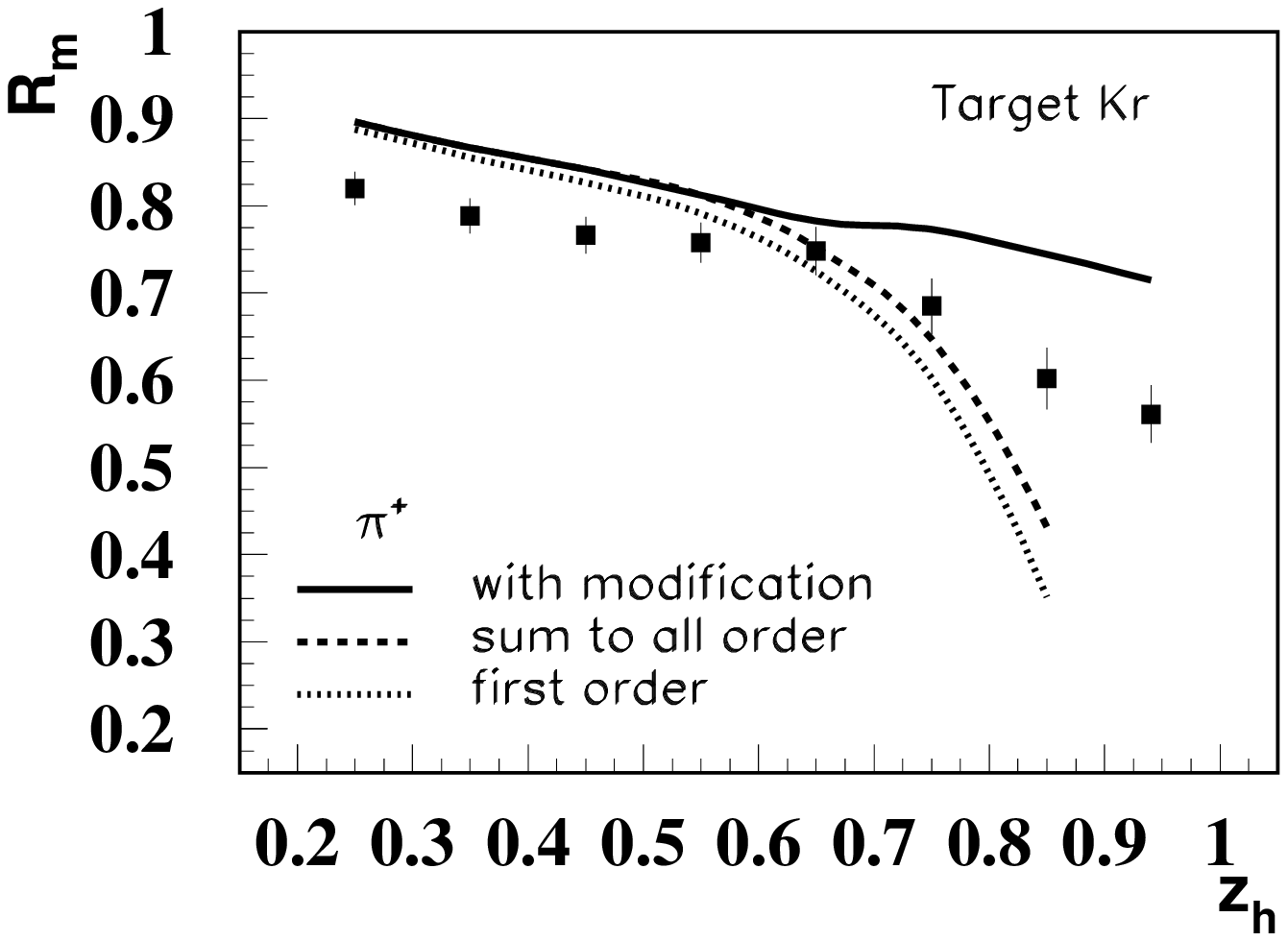}
\caption{Suppression from multiple scattering compared 
with HERMES data \protect\cite{hermes}.}
\label{fig2}
\end{center}
\end{figure}

\noindent{\it 4.\ Comparison with HERMES data}\
We compare our result with HERMES data \cite{hermes} in Fig.~\ref{fig2}.
$R_M$ is the ratio 
of hadron production rate rate per DIS event for a nuclear target 
$A$, $R^A$,  to that of a deuterium target $D$, $R^D$ \cite{guo-li},
\begin{equation}
R_M \equiv \frac{R^A}{R^D}
\approx 
\frac{\sum_q e_q^2\, \phi^A_q(x_B, Q^2)\, D_q(z_h+\Delta z,Q^2)}
     {\sum_q e_q^2\, \phi^D_q(x_B, Q^2)\, D_q(z_h,Q^2)}
\frac{\sum_q e_q^2\, \phi^D_q(x_B, Q^2)}
     {\sum_q e_q^2\, \phi^A_q(x_B, Q^2)} \ .
\label{RR}
\end{equation}
The dashed lines are from Eq.~(\ref{RR}) and the dot-dashed lines
includes only the double scattering.  
At large $z_h$, resummed theory curves are much
steeper than the data.  
This is because the formation time for the hadron or 
corresponding pre-hadron state is much shorter at large $z_h$, 
and therefore, the number of scattering for the propagating parton 
at large $z$, or the shift, should be reduced \cite{guo-li}.
Since the hadron formation time is proportional to 
$(1-z_h)$ \cite{alberto}, we modify
$\Delta z$ by multiplying a factor $(1-z_h)$ and 
produce the solid lines in Fig.~\ref{fig2}.

The suppression from the coherent multiple scattering, calculated here,
complements the suppression from medium-induced radiation 
\cite{induced-energylose}, which should give additional suppression and 
bring down the curves in Fig.~\ref{fig2} \cite{guo-li}.

\noindent{\it 5.\ Summary and outlook}\
 We showed that 
the coherent multiple scattering of a propagating quark 
in the medium, without the induced radiation, 
can change the quark fragmentation or 
hadron's production rate.  The net effect of 
leading power contributions in medium length 
is equivalent to a shift in the fragmentation 
function's $z_h \rightarrow z_h + \Delta z$, and
at the lowest order, $\Delta z$ is given by 
an universal matrix element.   
Our result  is complementary to the 
energy loss of induced radiation.  
However, beyond the leading order, the separation 
of the collisional energy loss and that of induced 
radiation will depend on the factorization scheme and 
is not unique, and needs further study.
Our approach can be generalized to hadron productions in $p+A$ 
and $A+A$ collisions.  

\end{document}